

\newbox\leftpage
\newdimen\fullhsize
\newdimen\hstitle
\newdimen\hsbody
\tolerance=1000\hfuzz=2pt

\def\bigans{b }
\def\answ{b }
\ifx\answ\bigans\message{(this will come out unreduced.}
\magnification=1200\baselineskip=20pt
\font\titlefnt=amr10 scaled\magstep3\global\let\absfnt=\tenrm
\font\titlemfnt=ammi10 scaled\magstep3\global\let\absmfnt=\teni
\font\titlesfnt=amsy10 scaled\magstep3\global\let\abssfnt=\tensy
\hsbody=\hsize \hstitle=\hsize 

\else\def\apans{h }
\message{(this will be reduced.}
\let\lr=l
\magnification=1000\baselineskip=16pt\voffset=-.31truein
\hstitle=8truein\hsbody=4.75truein\vsize=7truein\fullhsize=10truein
\ifx\apansw\apans\special{ps: landscape}\hoffset=-.54truein
  \else\hoffset=.05truein\fi
\font\titlefnt=amr10 scaled\magstep4 \font\absfnt=amr10 scaled\magstep1
\font\titlemfnt=ammi10 scaled \magstep4\font\absmfnt=ammi10 scaled\magstep1
\font\titlesfnt=amsy10 scaled \magstep4\font\abssfnt=amsy10 scaled\magstep1

\output={\ifnum\count0=1 
  \shipout\vbox{\hbox to \fullhsize{\hfill\pagebody\hfill}}\advancepageno
  \else
  \almostshipout{\leftline{\vbox{\pagebody\makefootline}}}\advancepageno
  \fi}
\def\almostshipout#1{\if l\lr \count1=1
      \global\setbox\leftpage=#1 \global\let\lr=r
   \else \count1=2
      \shipout\vbox{\hbox to\fullhsize{\box\leftpage\hfil#1}}
      \global\let\lr=l\fi}
\fi
\def\pmb#1{\setbox0=\hbox{#1}%
 \kern-.025em\copy0\kern-\wd0
 \kern .05em\copy0\kern-\wd0
 \kern-.025em\raise.0433em\box0 }

\def\title#1#2{\nopagenumbers\absfnt\hsize=\hstitle\rightline{}%
\centerline{\titlefnt\textfont0=\titlefnt%
\textfont1=\titlemfnt\textfont2=\titlesfnt #1}%
\centerline{\titlefnt\textfont0=\titlefnt%
\textfont1=\titlemfnt\textfont2=\titlesfnt #2
}%
\textfont0=\absfnt\textfont1=\absmfnt\textfont2=\abssfnt\vskip .5in}

\def\date#1{\vfill\leftline{#1}%
\tenrm\textfont0=\tenrm\textfont1=\teni\textfont2=\tensy%
\supereject\global\hsize=\hsbody%
\footline={\hss\tenrm\folio\hss}}
%

\def\nolabels{\def\eqnlabel##1{}\def\eqlabel##1{}\def\reflabel##1{}}
\def\writelabels{\def\eqnlabel##1{\hfill\rlap{\hskip.09in\string##1}}%
\def\eqlabel##1{\rlap{\hskip.09in\string##1}}%
\def\reflabel##1{\noexpand\llap{\string\string\string##1\hskip.31in}}}
\nolabels
%
\global\newcount\secno \global\secno=0
\global\newcount\meqno \global\meqno=1

\def\newsec#1{\global\advance\secno by1\xdef\secsym{\the\secno.}\global\meqno=1
\bigbreak\bigskip
\noindent{\bf\the\secno. #1}\par\nobreak\medskip\nobreak}
\xdef\secsym{}

\def\appendix#1#2{\global\meqno=1\xdef\secsym{#1.}\bigbreak\bigskip
\noindent{\bf Appendix #1. #2}\par\nobreak\medskip\nobreak}


\def\eqnn#1{\xdef #1{(\secsym\the\meqno)}%
\global\advance\meqno by1\eqnlabel#1}
\def\eqna#1{\xdef #1##1{(\secsym\the\meqno##1)}%
\global\advance\meqno by1\eqnlabel{#1$\{\}$}}
\def\eqn#1#2{\xdef #1{(\secsym\the\meqno)}\global\advance\meqno by1%
$$#2\eqno#1\eqlabel#1$$}

\global\newcount\ftno \global\ftno=1
\def\refsymbol{\ifcase\ftno
\or\dagger\or\ddagger\or\P\or\S\or\#\or @\or\ast\or\$\or\flat\or\natural
\or\sharp\or\forall
\or\oplus\or\ominus\or\otimes\or\oslash\or\amalg\or\diamond\or\triangle
\or a\or b \or c\or d\or e\or f\or g\or h\or i\or i\or j\or k\or l
\or m\or n\or p\or q\or s\or t\or u\or v\or w\or x \or y\or z\fi}
\def\foot#1{{\baselineskip=14pt\footnote{$^{\refsymbol}$}{#1}}\ %
\global\advance\ftno by1}


\global\newcount\refno \global\refno=1
\newwrite\rfile
\def\ref#1#2{$^{(\the\refno)}$\nref#1{#2}}
\def\nref#1#2{\xdef#1{$^{(\the\refno)}$}%
\ifnum\refno=1\immediate\openout\rfile=refs.tmp\fi%
\immediate\write\rfile{\noexpand\item{\the\refno.\ }\reflabel{#1}#2.}%
\global\advance\refno by1}
\def\addref#1{\immediate\write\rfile{\noexpand\item{}#1}}

\def\semi{;\hfil\noexpand\break}



\hyphenation{anom-aly anom-alies coun-ter-term coun-ter-terms}

\def\bR{{\bf R}}

\def\bJ{{\bf J}}

\def\bomg{\pmb{$\omega$}}

\centerline{{\bf Towards a theory of growing surfaces: Mapping
two-dimensional}}
\centerline{{\bf Laplacian growth onto Hamiltonian dynamics and statistics}}
\bigskip
\centerline{{\bf Raphael Blumenfeld}}

\centerline{Center for Nonlinear studies and Theoretical Division, MS B258}
\centerline{Los Alamos National Laboratory, Los Alamos, NM 87545, USA}

\bigskip
\item{}{\bf Abstract}
\smallskip
I show that the evolution of a two dimensional surface in a Laplacian field can
be described by Hamiltonian dynamics. First the growing region is mapped
conformally to the interior of the unit circle, creating in the process a set
of mathematical zeros and poles that evolve dynamically as the surface grows.
Then the dynamics of these quasi-particles is analysed. A class of arbitrary
initial conditions is discussed explicitly, where the surface-tension-free
Laplacian growth process is integrable. This formulation holds only as long as
the singularities of the map are confined to within the unit circle. But the
Hamiltonian structure further allows for surface tension to be introduced as an
energetic term that effects repulsion between the quasi-particles and the
surface. These results are used to formulate a first-principles statistical
theory of pattern formation in stochastic growth, where noise is a key player.

\bigskip

\date{LA-UR-93-3591}
\eject

\newsec{Introduction}
\bigskip

Growing surfaces in Laplacian fields are notoriously difficult to describe
theoretically in spite of the deceptive simplicity in formulating these
processes. A large body of examples exists, where the surfaces that result from
such growths exhibit a rich variety of patterns. Best-known are
diffusion-limited aggregation, solidification of supercooled liquid and
electrodeposition, to mention but a few.\ref\rev{For review see, e.g., P.
Pelce, "Dynamics of Curved Fronts" (Academic Press, San Diego, 1988); D. A.
Kessler, J. Koplik and H. Levine, Adv. Phys. {\bf 37}, 255 (1988); P. Meakin,
in "Phase Transitions and Critical Phenomena" Vol. 12 (Academic Press, 1988)
Eds. C. Domb and J. L. Lebowitz; T. Vicsek, "Fractal Growth Phenomena" (World
Scientific, Singapore, 1989)} Notwithstanding more than a decade of intensive
research into this problem and a large amount of literature there still exists
no theory that can predict, starting from the basic Eqs. of motion (EOM's), the
asymptotic patterns that such surfaces evolve into.\ref\piet{There are however,
probabilistic approaches where an iterative procedure is carried out for the
relevant probability density, or where a renormalization group approach is
used. See, e.g., L. Piteronero, A. Erzan and C. Evertsz, Phys. Rev. Lett. {\bf
61}, 861 (1988), Physica {\bf A151}, 207 (1988); T. C. Halsey and M. Leibig,
Phys. Rev. {\bf A 46}, 7793 (1992)}
Some time ago it has been proposed\ref\sb{B. Shraiman and D. Bensimon, Phys.
Rev. {\bf A 30}, 2840 (1984)} that in two dimensions conformal mapping be used
to transform the problem of growth of the physical surface to the dynamics of a
many-body system. The resulting Eqs. of motion of the relevant quasi-particles
(QP's) of the equivalent system are strongly coupled nonlinear first order
ODE's that are difficult to solve, other than for special cases.\ref\all{L.
Paterson, J. Fluid Mech. {\bf 113}, 513 (1981); L. Paterson, Phys. Fluids {\bf
28}, 26 (1985); S. D. Howison,  J. Fluid Mech. {\bf 167}, 439 (1986); D.
Bensimon and P. Pelce, Phys. Rev. {\bf A33}, 4477 (1986); S. Sarkar and M.
Jensen, Phys. Rev {\bf A35}, 1877 (1987); B. Derrida and V. Hakim, Phys. Rev.
{\bf A45}, 8759 (1992)} Unfortunately, this elegant formulation is doomed in
most cases to break down after a finite time due to the instability of the
surface with respect to growth of perturbations on ever short lengthscales,
which, in the absence of surface tension, develop into cusp singularities along
the physical surface. There have been some efforts to counteract this breakdown
by using surface tension to cut off the short lengthscales in a renormalizable
manner,\ref\leo{D. Bensimon, L. P. Kadanoff, S. Liang, B. I. Shraiman and C.
Tang, Rev. Mod. Phys. {\bf 58}, 977 (1986); W-s Dai, L. P. Kadanoff and S.
Zhou, Phys. Rev. {\bf A 43}, 6672 (1991)} but these approaches are somewhat
ad-hoc in that the procedure for cutting off the short scales can be
arbitrarily chosen. Another approach has been proposed recently to prevent
formation of such cusps in which tip-splitting acts as a mechanism to reduce
the locally high surface curvature energy.\ref\bbi{R. Blumenfeld and R. C.
Ball, in preparation} I will propose here a new approach to deal with this
problem.

Another issue in this context that is deemed important for various reaseons is
whether the system is integrable, or even more weakly, can it be described by a
Hamiltonian structure. It has been already suggested that the problem enjoys a
set of conserved quantities,\ref\mm{S. Richardson, J. Fluid Mech., {\bf 56},
609 (1972); M. B. Mineev, Physica {\bf D 43}, 288 (1990)} but so far it has
been unclear whether these quantities can assist in finding an energy-like
functional in the problem.

The main issues addressed in this paper are: 1) It is shown that the
surface-tension-free surface follows Hamiltonian dynamics, once described with
the proper variables. The Hamiltonian can be chosen to be of separable
action-angle variables which describe then an orbital motion on a
$2N$-dimensional torus with $2N$ being half the number of the degrees of
freedom. This transformation is acheived  by first writing down explicitly the
EOM's in the equivalent many-body system and then transforming the coordinates
of the QP's into a set of action-angle variables. The general set of equations
that effect this transformation is given explicitly. 2) I analyse a specific
system as an example and demonstrate its integrability explicitly. 3)  Using
the Hamiltonian formalism I propose to introduce surface energy simply as a
term in the Hamiltonian, which acts as an effective repulsive potential between
the QP's and the surface and rejects cusp formation and a subsequent breakdown
of the formalism. 4) I discuss how to use these results and formulate the steps
to construct a {\it first-principles theory} for pattern formation in
stochastic Laplacian growths.

\smallskip
\newsec{Formulation of the problem and Hamiltonian dynamics}
\bigskip

The problem under study can be formulated as follows: Consider a simply
connected line surface, $\gamma(s)$, embedded in two dimensions that is
parametrized by $0\leq s <2\pi$, and which is fixed at a given electrostatic
potential (or concentration for diffusion controlled processes). A higher
potential is assigned to a very far circular boundary. The field, $\Phi$,
outside the area enclosed by $\gamma$ is determined by Laplace's Eq.
\eqn\L{\nabla^2 \Phi = 0\ .}
In addition the surface is assumed to grow at a rate that is proportional to
the local gradient of the field normal to the surface
$$v_n = - {\bf \nabla}\Phi\cdot\hat{\bf n}\ .$$

It has been shown\sb that the surface evolves in time, $t$, according to
\eqn\Ai{\partial_t\gamma(s,t) = -i\partial_s\gamma(s,t)
\Bigl[|\partial_s\gamma(s,t)|^{-2} +
ig\Bigl(|\partial_s\gamma(s,t)|^{-2}\Bigr)\Bigr]\ .}
This equation is the limit of a conformal map $\zeta=F(z,t)$ that maps the UC
onto the physical surface in the $\zeta$-plane through $\gamma(s,t)=\lim_{z\to
e^{is}} F(z,t)$. The map that I consider here is quite general and is defined
as the ratio of two polynomials of the same degree. This map preserves the
topology far away from the growth and therefore the original boundary
conditions remain unchanged\bbi
\eqn\Aii{ {{dF(z)}\over{dz}} = \prod_{n=1}^N {{z-Z_n}\over{z-P_n}}\ ,}
where $\{Z\}$ and $\{P\}$ are the zeros and the poles of the map, respectively.
It can be shown\sb$^,$\bbi that the dynamics of these singularities are
governed by the EOM's
\eqn\Aiii{\eqalign{-\dot Z_n &= Z_n\Bigl\{G_0 + \sum_{m'} {{Q_n +
Q_{m'}}\over{Z_n - Z_{m'}}} \Bigr\} + Q_n\Bigl\{1 - \sum_m{{Z_n}\over{Z_n -
P_m}}\Bigr\} \equiv f_n(\{Z\};\{P\}) \cr
-\dot P_n &= P_n\Bigl\{G_0 + \sum_m {{Q_m}\over{P_n - Z_m}}\Bigr\} \equiv
g_n(\{Z\};\{P\})\ , \cr }}
where
$$\eqalign{Q_n &= 2\prod_{m=1}^N\Bigl[ {{(1/Z_n - P_m^*)(Z_n-P_m)}\over
{(1/Z_n - Z_m^*)(Z_n - Z_{m'})}}\Bigr] \cr
G_0 &= \sum_{m=1}^N {{Q_m}\over{2 Z_m}} + \prod_{m=1}^N {{P_m}\over{Z_m}}\ ,
\cr}$$
and where the primed index indicates $m'\neq n$.

The fact that the surface evolves via Hamiltonian dynamics can be hinted upon
already from Eq. $\Ai$. Writing only the evolution of the normal component and
ignoring the physically insignificant 'sliding' of a point along the surface,
we have
\eqn\Hi{{{\partial\gamma}\over{\partial t}} = -i{{\delta s}\over{\delta
\gamma^*}}\ ,}
where $^*$ stands for complex conjugate. This form has a Hamiltonian structure
if $s$ is interpreted as Hamiltonian and $\gamma$ and $\gamma^*$ are the
conjugate fields. Alternativelky a Hamiltonian structure can be observed also
in the evolution of the conformal map. The generalized form of Eq. $\Ai$ is
\eqn\ARi{{{\partial F'}\over{\partial t}} = {{\partial}\over{\partial z}}
\Bigl[ z F' G\Bigr]\ ,}
where $F'\equiv dF/dz$ and where $G$ is the analytic function in the square
brackets on the r.h.s. of $\Ai$. A straightforward manipulation leads to
\eqn\Arii{{{\partial z}\over{\partial t}} = -{{\partial}\over{\partial
F'}}\Bigl[z F' G\Bigr] \ .}
Therefore we can identify $z$ and $F'$ as the conjugate variables of the
Hamiltonian function
\eqn\ARii{\tilde H = z F' G\ .}
Although this particular derivation does not immediately insure integrability,
it establishes that the surface growth is governed by Hamiltonian dynamics,
which is the most important issue for the purpose of this presentation.

The Hamilton form into which we wish to map the dynamical system,
$$H=H(\{J\};\{\Theta\})\ ,$$
is a function of new cannonical action-angle two-dimensional (or complex)
variables that depend on the coordinates $\{Z\}$ and $\{P\}$. Moreover, we
require that $H=H\bigl(\{J\}\bigr)$ because we expect that there are
sufficiently many relevant arbitrary constants of the motion as has already
been found,\mm$^,$\ref\bbii{R. Blumenfeld, preprint} and consequently that the
motion is on a $2N$-dimensional torus (there are $2N$ QP's each of which is
specified by two spatial coordinates, and therefore altogether $4N$ degrees of
freedom). From previous results\bbii it appears that the chosen Hamiltonian is
not unique and we can therefore choose the simple separable form
\eqn\Av{H=\sum_{n=1}^N \omega_n J_n\ .}
It should be emphasized that the fact that the Hamiltonian is separable is of
no consequence to the validity of the present formulation, but it clearly
demonstrates the integrability of the problem. The action-angle variables are
required to obey Hamilton-Jacobi EOM's
\eqn\Avi{\dot\Theta_n = {{\partial H}\over{\partial J_n}}\ \ \ ;\ \ \ \dot{J_n}
= -{{\partial H}\over{\partial \Theta_n}}\ ,}
which for the Hamiltonian chosen here reduce to the traditional simple
constants $\omega_n$ and $N$ zeros. We again stress that all these variables
are two-dimensional vectors.

We seek now a transformation that gives $J_n=J_n(\{Z\};\{P\})$ and
$\Theta_n=\Theta_n(\{Z\};\{P\})$. The general set of Eqs. that facilitate such
a transformation is found by combining Eqs. $\Aiii$, $\Av$ and $\Avi$
\eqn\Avii{\dot J_n = \sum_{m=1}^N \Bigl[{{\partial J_n}\over{\partial Z_m}}f_m
+ {{\partial J_n}\over{\partial P_m}}g_m \Bigr] = -{{\partial H}\over{\partial
\Theta_n}} = 0 }
\eqn\Aviii{\dot \Theta_n = \sum_{m=1}^N \Bigl[{{\partial
\Theta_n}\over{\partial Z_m}}f_m + {{\partial \Theta_n}\over{\partial P_m}}g_m
\Bigr] = {{\partial H}\over{\partial J_n}} = \omega_n \ ,}
(the r.h.s. of these Eqs. is particular to the Hamiltonian that has been chosen
in Eq. $\Av$). The set of equations $\Avii$ and $\Aviii$ can be written in a
more general form:
\eqn\Adi{\bigl({\bf f}(\bR)\cdot{\bf \nabla}\bigr) \bJ(\bR) = \bomg \ .}
In this notation ${\bf f}$, $\bJ$, $\bR$ and $\bomg$ are $2N$-components
(complex) vectors
$$\eqalign{ {\bf f} &\equiv \bigl(f_1, f_2,\dots, f_N, g_1, \dots, g_N \bigr) \
\ ; \ \ \ \bR \equiv \bigl(Z_1, Z_2,\dots, Z_N, P_1, \dots, P_N \bigr) \cr
{\bJ} &\equiv \bigl(J_1, J_2,\dots, J_N, \Theta_1, \dots, \Theta_N \bigr) \ \ ;
\ \bomg \equiv \bigl(-{{\partial H}\over{\partial \Theta_1}}, -{{\partial
H}\over{\partial \Theta_2}},\dots, -{{\partial H}\over{\partial \Theta_N}},
{{\partial H}\over{\partial J_1}}, \dots, {{\partial H}\over{\partial J_N}}
\bigr) \ , \cr}$$
and ${\bf \nabla}$ is the gradient in the $2N$-dimensional space of $\bR$. If
$\bomg$ is a vector of $N$ complex zeros and $N$ complex constants of the
motion, as chosen in Eq. $\Av$ then $\Adi$ is a linear set and therefore has a
solution as long as the known operator ${\bf f}(\bR)\cdot{\bf \nabla}$ has no
vanishing eigenvalue. Thus, unless such a singular case occurs, this set of
Eqs. must have a solution for $\{\bJ(\bR)\}$. This solution defines a specific
transformation from the chosen Hamiltonian to the problem of the dynamics of
the singularities, and hence to the original growth problem. The existence of
such a solution immediately points to the integrability of the system.

\smallskip
\newsec{A case study: N-symmetric growth}
\bigskip

Having substantiated the existence of an underlying Hamiltonian structure
having formulated the general transformation, let us consider a specific
example of a growth where the solution to the set of Eqs. $\Avii$ and $\Aviii$
can be discussed explicitly. This will illustrate the formalism and show a
particular case where the system is indeed integrable. We represent the initial
surface at $t=0$ by the form
\eqn\Bi{\gamma(s,0) = e^{is} + \sum_{n=1}^N R_n\ln\bigl(e^{is}-P_n(0)\bigr)\ ,}
where we have defined the quantities $R_n = \prod_{m'\neq n}^N
\bigl[\bigl(P_n(0) - Z_{m'}(0)\bigr)/\bigl(P_n(0)-P_{m'}(0)\bigr)\bigr]$. The
surface $\gamma(s,t)$ can be shown to be describable by this form for any later
time by simply substituting for $P_n$ and $Z_n$ their time-dependent values.
The form $\Bi$ remains valid for any number of singularities when $R_n$ are
interpretd as the residues of the analytic map $dF/dz$. Moreover, it can be
shown\sb that the number of singularities, $N$, of each kind is invariant under
the EOM's. The growth problem consists now of finding the dynamics of $N$ zeros
at $Z_n(t)=Z(t)e^{in\alpha}$, where $\alpha\equiv 2\pi/N$, and $N$ poles at
$P_n(t)=P(t)e^{in\alpha}$ ($Z(t)$ and $P(t)$ are real). The trajectories of
these singularities can be found by substituting directly into the EOM's
$\Aiii$. It is worthwhile first to note that both from symmetry arguments and
from direct analysis of the EOM's it can be shown that the motion of all the
singularities will be purely radial. Therefore the only relevant quantities to
analyse are the distance of a zero, $Z(t)$, and a pole, $P(t)$, from the
origin. It can also be shown\bbi$^,$\bbii that the analyticity requirement
imposes a sum rule on the locations of the singularities
\eqn\Biii{\sum_{n=1}^N Z_n(t) = \sum_{n=1}^N P_n(t)\ ,}
which is identical to requiring that
$$\sum_{n=1}^N R_n = 0\ .$$

All these constraints immediately simplify the EOM's and these can now be cast
in the form
\eqn\Biiia{\eqalign{-{1\over {Nx}} {\dot x} &= y/x - K \bigl[1 - 2/N + (1 +
2/N)y/x \bigr] \cr
-{1\over {Ny}} {\dot y} &= y/x - K (1 + y/x) \ ,\cr} }
where $x\equiv Z^N ,\ y\equiv P^N$ and $K\equiv (1-xy)/(1-x^2)$. The behaviour
of the system differs when $P(0)$ is smaller or larger than $Z(0)$. In the
first case cusp singularities appear at a finite time, while in the second such
singularities may be avoided under appropriately chosen initial conditions. The
issue of cuspless growths due to particular choices of initial conditions has
been discussed recently,\ref\kad{cuspless growths: S.D. Howison, SIAM J. Appl.
Math. {\bf 46}, 20 (1986); D. Bensimon and P. Pelce, Phys. Rev. {\bf A 33},
4477 (1986); M. Mineev-Weinstein and S.P. Dawson, preprint (1993)} and is not
very relevant to the main thrust of this paper. I will only remark in passing
that recent calculations\ref\biii{R. Blumenfeld, unpublished} for the present
N-symmetric case suggest that the non-cuspiodal solution for $P(0)>Z(0)$ is
unstable and may, under perturbation, flow into a cusp-forming solution. Here I
choose for initial conditions $P(0)<Z(0)$ and arg$(P_n)={\rm arg}(Z_n)$, where
the solution can be shown to break down after a finite time. From the EOM's it
can be seen that if the zero and the pole meet, say at $r_0$, they keep moving
together at an exponential rate,
$$r(t) = r_0 e^{Nt}\ .$$
This calculation, however, is academic because once a pole and a zero meet they
'anihilate' and disappear from the scene, as can be verified by inspecting the
form of the conformal map $\Aii$.

Another comment is in place now: The fact that the solution of the rpoblem
under study here may not be valid for $t\to\infty$ is of no concern for the
thrust of the present formulation: First because at this stage no claim is made
regarding the asymptotic structure of the surface, and second, the aim at this
point is to facilitate a mapping of the growth problem onto Hamitonian
dynamics, even if this map is relevant only for a finite time, during which the
EOM's $\Aiii$ are valid. Indeed, one of the present goals, which is addressed
in more detail below, is to extend this regime of validity by introducing a
mechanism that prevents cuspiodal solutions with the aid of surface tension.

Using $\Bi$ and a back-of-an-envelope algebraic manipulation gives for the
explicit form of the interface at any time $t>0$:
\eqn\Biiiz{\gamma(s,t) = e^{is} - {{Z(t) (1 - y/x)}\over N} \sum_{n=1}^N
e^{2n\pi i/N} \ln\bigl(e^{is} - P(t) e^{2n\pi i/N}\bigr)\ .}

A rather long and cumbersome manipulation of the EOM's $\Biiia$, which will be
presented explicitly elsewhere, yields that the following is a constant of the
motion
\eqn\Biiic{{{y^{1-2/N}}\over{x-y}} + F(y) = Const\ ,}
where
$$F(y) \equiv {1\over N} \int^{y^2} {{u^{-1/N} du}\over{1-u}}\ .$$
The EOM's can be integrated out now and the motion of $P$ and $Z$ can be found
explicitly. Several stages of the growth are shown in the Fig.

It should be noted that in the above treatment I have considered a unity factor
in front of the conformal map $F$, while in fact the map should be multiplied
by a purely time-dependent term, $A(t)$. Since an implicit assumption in the
formulation of the problem, as given above, is that the flux into the growth is
constant in time then the total area enclosed by the surface should increase
linearly with time. It follows that by maintaining the unit prefactor the
growth is practically rescaled at each time step by $1/A(t)$. This is the
reason why a close inspection of the Fig. can reveal sections of the boundary
that seem to retreat with time, although the actual physical surface always
grows outwards. The evolution of this prefactor follows a first order ordinary
differential Eq. as has already been discussed in the literature for other
forms of the map\all$^,$\leo and it can be very simply incorporated into the
formulation, but for clarity this is ommitted here.
\smallskip
Eqs. $\Avii$ and $\Aviii$ are now used to find the action-angle variables in
terms of the original coordinates. We can immediately set the value of the
action variable to
\eqn\Di{J = {{y^{1-2/N}}\over{x-y}} + F(y)\ ,}
which we know is a constant of the motion. The Hamiltonian is then
$$H = \omega J\ ,$$
with $\omega$ a constant that can be found by using Eqs. $\Avii$ and $\Aviii$,
and where $\Theta = \omega t + \Theta_0$. The fact that we have only one action
and one angle variables reflects the degeneracy of the problem due to the
$N$-rotational symmetry, where only $Z(t)$ and $P(t)$ remain the relevant
degrees of freedom.

\smallskip
\newsec{Introduction of surface tension}
\bigskip

Let us now consider the role that surface tension plays in this picture. It is
evident that in real growth processes cusps do not form either because
microscopic atomistic mechanisms (which are completely smoothened over in the
continuum description) become relevant, or because there is a macroscopic
surface energy to be paid when the curvature of the surface increases. Since
the entire formulation above is given in the continuum we are outside the
microscopic regime, and I therefore consider only the latter effect. If we try
to imagine what the actual effect of surface energy would translate into in the
many-body picture, we can immediately see that this would correspond to an {\it
effective repulsion} in the mathematical plane between a QP and the internal
boundary defined by the UC. It is important to emphasize that only because a
Hamiltonian formulation is now available the term 'repulsion' can have a
physical meaning. Previously this effect could only be incorporated in the
EOM's as an additional ad-hoc term.
Now, however, having a Hamiltonian at hand, we can immediately translate the
surface energy in the physical plane into a repulsive potential term that
enters the Hamiltonian in the mathematical plane (An equivalent and maybe even
more natural approach is to identify the physical constant quantity that this
Hamiltonian corresponds to in the physical plane when possible. For example,
this could be simply related to the area of the UC or its circumference).
Repelling the QP the surface avoids forming a cusp, by simply not letting the
QP approach too close. Thus we can now consider the many body problem as QP's
confined within a potential well that consists of a circular infinite wall. A
simpler approach with this idea in mind has been considered by Blumenfeld and
Ball,\bbi where such an effect was introduced (albeit, in an ad-hoc manner
through the EOM's and not in a Hamiltonian form), which initiated creation of
particles and anti-particles (poles and zeros). This resembles very much a
field formulation, where particles can be spontaneously created from vacuum
fluctuations, as well as annihilated when anti-particles meet, as discussed
above. In the present case I argue that the simplest term to consider as such a
repulsive potential, $V$, in the {\it mathematical plane} is
\eqn\Dii{{\rm Re} V = \sigma \lim_{z\to e^{is}} \Bigl[\mid F'\mid
K(\{Z\},\{P\})\Bigr]\ ,}
where $K$ is the curvature in terms of the locations of the zeros and the
poles\bbi
\eqn\Diii{K(\{Z\},\{P\}) = \lim_{z\to e^{is}} \mid F'\mid^{-1} \Bigl\{ 1 + {\rm
Re}\sum_{n=1}^N\bigl\{ {z\over{z-Z_n}} - {z\over{z-P_n}} \bigr\} \Bigr\}\ .}
The imaginary part of the $V$ is found by imposing analyticity (or requiring
that Cauchy-Riemann relations hold) outside the growth
\eqn\Diiii{V = {1\over{2 \pi i}}\lim_{\epsilon\to 0} \oint {{z + z'}\over{z +
\epsilon - z'}} {\rm Re}\left\{V\right\}
{{d z'}\over{z'}}\ .}
This potential term in the Hamiltonian is surprisingly easy to handle in the
mathematical plane in term of zeros and poles because it decouples naturally to
a sum of individual contributions of the QP's, namely
\eqn\Div{\eqalign{{\rm Re}V_0 &= \sigma \cr
{\rm Re} V(Z_n) &= \sigma {\rm Re} {z\over{z-Z_n}} \cr
{\rm Re}V(P_n) &= -\sigma {\rm Re} {z\over{z-P_n}}\ . \cr}}
This unexpected decoupling should lead to a considerable simplification of the
dynamics. But it should also be recalled that the canonical variables we are
now working with are the action-angle variables rather than the zeros and poles
locations. The transformation between the variables is directly related to the
quantities given in Eq. $\Adi$ and the generalized force in the $m$-th
direction is given by
\eqn\Dv{ -\{\nabla_J V\}_m \equiv -{{\partial V}\over{\partial J_m}} = -\bigl\{
{\bf {\hat A}}^{-1} {\bf b} \bigr\}_m\ ,}
where
$${\bf {\hat A}}_{jk}\equiv {{\partial J_j}\over{\partial R_k}}$$
can be found from Eqs. $\Avii$-$\Adi$,
$${\bf b}_j \equiv {{\partial V}\over{\partial R_j}}\ ,$$
and where use of the orthogonality of the action variables has been made in
$\Dv$.

\smallskip
\newsec{Effects of noise and a statistical formulation of the theory}
\bigskip

The next, and probably the most difficult, step towards a theory of growth is
to include the effect of the noise. As is well known in Laplacian growths, the
patterns that such processes evolve into depend very crucially on the nature of
the noise in the system. This noise can originate from many sources: general
fluctuations in the local Laplacian field, discretization of the underlying
background over which the field is solved (lattice growth), discretization of
the incoming flux in the form of finite size particles that stick to the
growing aggregate (e.g., diffusion-limited-aggregation and similar processes),
etc.. In the present formulation the effects of all those can be interpreted as
simply 'smearing' the predetermined trajectories of the QP's in the
mathematical plane. However, now we have a Hamiltonian available, which
immediately points to the existence of a Liouville's theorem, namely, that the
distribution of the canonical variables in phase space is incompressible. Thus
it is straightforward to write down an EOM for the time evolution of this
distribution and consequently it can be possible to analyse its asymptotic
behaviour. This exercise is currently being carried out by this author and will
be reported at a later time. Either from such a calculation, or via
phenomenological argument, one can now devise a measure
$\mu\bigl(H(\{\bJ\})\bigr)$ (e.g., the Gibbs measure, $e^{-\beta H}$) and to
calculate {\it average} quantities weighted by this measure
\eqn\Dvi{\langle X \rangle = {1\over \cZ} \int X\  \mu\bigl(H(\{J\})\bigr)
d^{N}J
\ ,}
where the partition function $\cZ$ is
$$\cZ \equiv \int \mu\bigl(H(\{J\})\bigr) d^{N}J\ .$$
Suppose that the Gibbs measure is indeed the relevant measure for this purpose.
Then the Lagrange multiplier, $\beta$, which in traditional statistical
mechanics is associated with the temperature, would correspond here to the
effective magnitude of the noise. This issue is also under current
investigation.

Using this formalism we can now describe in a well defined way the statistics
of the QP's and in general any properties that depend explicitly on the
distribution of their locations. This is is possible because by the above
arguments this distribution must flow to a stable limiting form. Indeed, it is
well known that for many growth processes in Laplacian (and in other) fields
the growth probability along the surface seems to flow towards a stable
asynptotic form. One manifestation of this is the appearance of a
time-independent multifractal function.\ref\mf{B. B. Mandelbrot, J. Fluid Mech,
${\bf 62}$, 331 (1974); Ann. Israel Phys. Soc. ${\bf 2}$, 225 (1978); T. C.
Halsey, M. H. Jensen, L. P. Kadanoff, I. Procaccia and B. I. Shraiman, Phys.
Rev. A${\bf 33}$, 1141 (1986); T. C. Halsey, P. Meakin, and I. Procaccia, Phys.
Rev. Lett. ${\bf 56}$, 854 (1986); C. Amitrano, A. Coniglio, and F. diLiberto,
Phys. Rev. Lett.
${\bf 57}$, 1016 (1986); P. Meakin, in {\it {Phase \ Transitions \ and \
Critical \ Phenomena}} Vol. ${\bf 12}$, edited by C. Domb and J.L. Lebowitz
(Academic Press, New York 1988), p.335; R. Blumenfeld and A. Aharony, Phys.
Rev. Lett. ${\bf 62}$,  2977 (1989)} Since it is possible to show that the
growth probability along the growing surface is directly related to the spatial
distribution of $\{Z\}$ and $\{P\}$,\ref\biv{R. Blumenfeld, in preparation} one
can therefore analytically predict the statistics of the physical surface, its
{\it asymptotic morphology}, and in particular the entire multifractal
spectrum. Thus, it is this author's belief that this statistical formulation of
the problem of a growing free surface in a Laplacian field is a nucleus of a
full theory of growth.

\smallskip
\newsec{Discussion and concluding remarks}
\bigskip

To conclude I have shown here that the growth of a surface in a Laplacian field
is governed by Hamiltonian dynamics. I have chosen the simple decoupled
Hamiltonian given in $\Av$ to illustrate how this transformation can be carried
out, and have formulated the transformation Eqs. for a general surface. I have
also shown that for one particular example such a transformation can be derived
explicitly and the resulting Hamiltonian leads to an integrable system. The
question whether integrability exists in the general case for arbitrarily
chosen initial surface depends on whether there is at least one solution to Eq.
$\Adi$ and the condition for this has been given explicitly. Regarding this
point several questions still loom: i) Can the operator  ${\bf f}(\bR)\cdot{\bf
\nabla}$ have a vanishing eigenvalue? if so what is the physical surface that
this situation corresponds to? ii) Can we choose any Hamiltonian onto which to
map the system or are we constrained in this choice?

Next I have proposed to interpret surface energy as giving rise to a repulsive
potential term in the Hamiltonian and I have derived the force which
corresponds to a repulsion between the QP's and the surface. This repulsion
seems the most natural way to introduce surface tension to prevent cusp
singularities from forming along the surface due to the inherent instability of
thie growth process.

Up to this point only deterministic trajectories of the system have been
discussed. The next step towards constructing a theory of growth consists of
including the noise that is fundamental to the stochastic process by taking
into account its effect on the deterministic trajectories of the QP's.
In the case of Gibbs measure, $\mu(H) = e^{-\beta H}$, this can be done by
identifying the Lagrange multiplier, $\beta$, that corresponds to the
'smearing' of the trajectory due to fluctuations in the environment. This
introduction of noise is analogous to (but seems to this author somewhat more
natural than) introducing noise effects directly in the EOM's of the QP's. The
equivalence can be probably demonstrated through an analogue of the
fluctuation-dissipation theorem, which has not been derived yet in this
context. From this it is straightforward to construct a partition function,
$\cZ$, from which all quantities that depend on the noise variable $\beta$ and
the distribution of QP's can be derived. The last and final step is probably
the easiest technically and consists of relating the morphology of the evolving
surface to the distribution of QP's in space, thus enabling the translation of
the averages defined above into the statistics of the surface. It is this
author's belief that in this way the fractal dimension of the surface can be
derived, as well as the entire so-called multifractal function. This belief is
based on the fact that these quantities relate only to the growth probability
(or the normal growth rate) along the surface. The growth probability relates
simply to the 'charge' (in the context of electrodeposition) along the surface
which depends only on the curvature. But the curvature, in turn, is expressible
in terms of only the locations of the QP's. It follows that if the distribution
of the latter flows towards a stable limit, so does the asymptotic statistics
of the surface and the above formalism should be able to predict all the
quantities that are relevant to the morphology. Of course, much of the results
still depend on the kind of noise that 'smears' the initially deterministic
trajectories, but this information can be considered either as a physical input
or it may be possible to derive from the master equation formulation.

As a final remark let me mention that investigation is being directed currently
towards extending this formalism towards growth of surfaces in higher
dimensions, as well as in general non-Laplacian fields.

\bigskip
{\bf Acknowledgement}

I thank G. Berman for illuminating discussions on the hamiltonian form $\Av$
and to M. Mineev-Weinstein for discussions on the possible integrability of
Laplacian growth.

\bigskip
{\bf FIGURE CAPTION}
\bigskip
\item{1.} The growth of the surface exemplified in the text for two zeros and
two poles. Half the surface is shown. Inset: the cusp formation of the tip.
\vfill\eject\immediate\closeout\rfile
\baselineskip=18pt\centerline{{\bf REFERENCES}}\bigskip\frenchspacing%
\input refs.tmp\vfill\eject\nonfrenchspacing
\bye